\documentclass[%
 reprint,
superscriptaddress,
showpacs,preprintnumbers,
 amsmath,amssymb,
 aps,
 prl,
twocolumn,
prstab,
nofootinbib
]{revtex4-1}

\usepackage{scrextend}
\usepackage{graphicx}
\usepackage{dcolumn}
\usepackage{bm}
\usepackage{amsmath}
\usepackage[colorlinks=true]{hyperref}

\newcommand{\abs}[1]{\left\vert#1\right\vert}

\begin{document}
\title{Computation and Numerical Simulation of Focused Undulator Radiation\\
	for Optical Stochastic Cooling}
\author{M.B. Andorf}
\affiliation{Northern Illinois Center for Accelerator \& Detector Development and Department of Physics, Northern Illinois University, DeKalb IL 60115, USA} 
\author{V.A. Lebedev}
\affiliation{Fermi National Accelerator Laboratory, Batavia, IL 60510, USA}
\author{J. Jarvis}
\affiliation{Fermi National Accelerator Laboratory, Batavia, IL 60510, USA}
\author{P. Piot}
\affiliation{Northern Illinois Center for Accelerator \& Detector Development and Department of Physics, Northern Illinois University, DeKalb IL 60115, USA} 
\affiliation{Fermi National Accelerator Laboratory, Batavia, IL 60510, USA}

\begin{abstract}
Optical stochastic cooling (OSC) is a promising technique for the cooling of dense particle beams. Its operation at optical frequencies enables obtaining a much larger bandwidth compared to the well-known microwave-based stochastic cooling. In the OSC undulator radiation generated by a particle in an upstream ``pickup" undulator is amplified and focused at the location of a downstream ''kicker" undulator. Inside the kicker, a particle interacts with its own radiation field from the pickup. The resulting interaction produces a longitudinal kick with its value depending on the particles momentum which, when correctly phased, yields to longitudinal cooling. The horizontal cooling is achieved by introducing a coupling between longitudinal and horizontal degrees of freedom. Vertical cooling is achieved by coupling between horizontal and vertical motions in the ring. In this paper, we present formulae for computation of the corrective kick and validate them against numerical simulations performed using a wave-optics computer program.
\end{abstract}
	

\maketitle

\section{Introduction}
Optical Stochastic Cooling (OSC) is a method of beam cooling in an accelerator that utilizes the short, radiation wave-packet generated by a particle passing through an undulator, as a way to make a corrective kick in energy with minimal interference from other nearby particles in the beam. The method is an extension of the well-known microwave stochastic cooling~\cite{Mohl,vdmeer,Bramham} to optical wavelengths. A transition from microwave to optical frequencies enables an increase of the cooling bandwidth by 3-4 orders of magnitude. For a system operating at the optimal gain that shortens the cooling time in the same proportion~\cite{OSC_Mikhailichenko,OSC_Zolotorev}.\\

\begin{figure}[bbb!!!]
	\centering
	\includegraphics*[width=0.485\textwidth]{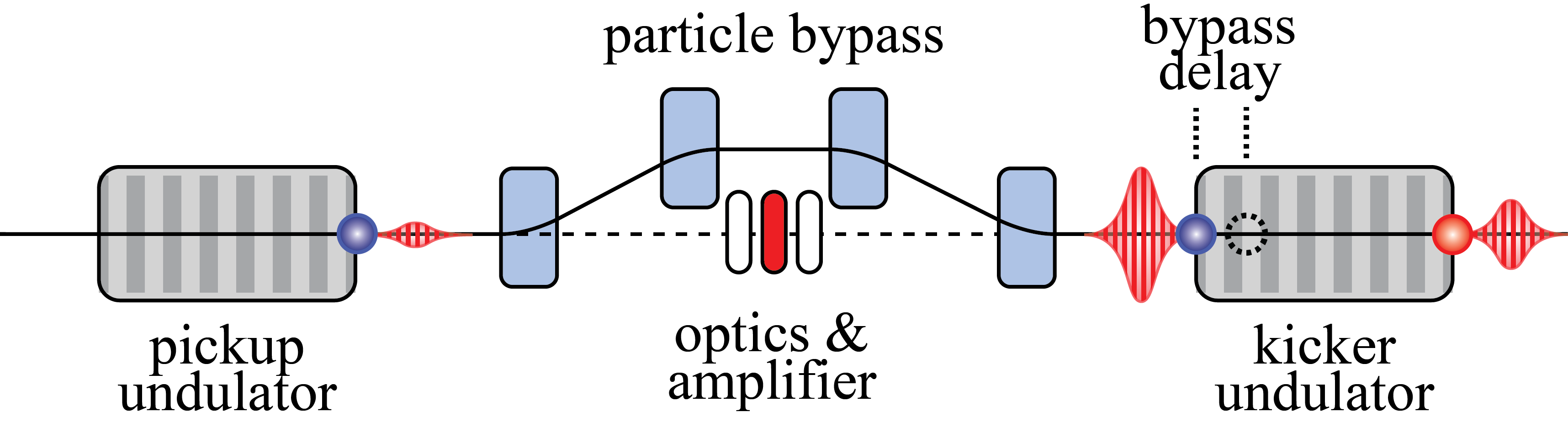}
	\caption{Conceptual diagram of a transit-time OSC insertion. The blue circles and red wiggles respectively represent the charged-particle beam and laser pulse. The beam passes through a by-pass beamline made of four dipole magnets (blue rectangle) and other (not shown) focusing elements. }
	\label{fig_OSC_schem}
\end{figure}
In OSC the corrective kick occurs via interaction of the particle with its own radiation, radiated at an earlier time in a ``pickup" undulator,  imaged in a downstream identical undulator referred to as the ``kicker"; See Fig.~\ref{fig_OSC_schem}. The amount of energy exchange between the particle and its radiation is determined by the arrival time of the particle with respect to the radiation wave-packet. This difference in arrival time depends on the particles phase-space coordinates and the lattice optics associated with the bypass beamline (typically a magnetic chicane) inserted between the pickup and kicker. For a properly tuned chicane averaging kicks over many turns simultaneously provides cooling in the horizontal and longitudinal planes. Although the kick is corrective in energy only; the coupling between longitudinal and horizontal motions introduced by the beamline between pickup and kicker provides cooling in the horizontal plane. The coupling is achieved by correct choice of pickup-to-kicker transfer matrix and the introduction of dispersion in both undulators. 

Despite its great promise to help store high intensity beams~\cite{blaskiewicz}, and cool different types of beams~\cite{babzien,muon,tevatron}, OSC has so far not been experimentally demonstrated. A possible proof-of-principle experiment is in preparation at Fermilab~\cite{IOTAVal}. The experiment would be carried out using a 100-MeV electron beam stored in the integrable-optics test accelerator (IOTA), a small-circumference ring dedicated to address issues pertaining to high-intensity beams. \\

\begin{figure}[ttt!!!!!!!!!]
	\centering
	\includegraphics*[width=0.485\textwidth]{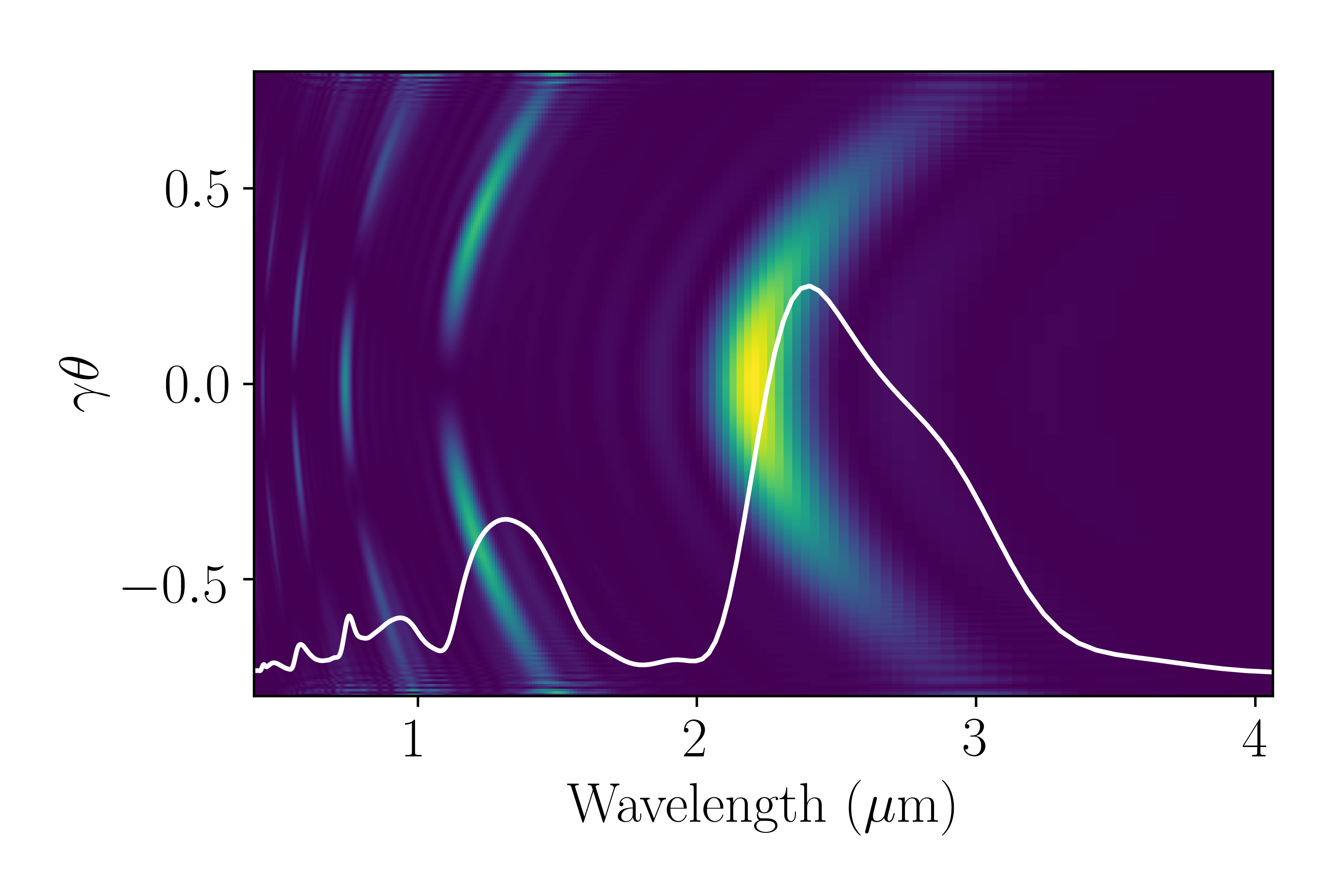}
	\caption{Undulator radiation spectrum in the horizontal plane computed with SRW assuming the parameters given in Table~\ref{und_params}. The white trace integrates the spectrum in the transverse plane over an angular acceptance corresponding to $\gamma\theta_m=0.8$.}
	\label{fig:undrad}
\end{figure}

In this paper, we present a semi-analytical theory to compute the energy kick which a particle receives passing through an OSC system.  We apply this theory to the 
proof-of-principle test of the OSC to be carried out in the IOTA ring at Fermilab~\cite{IOTA}. We should however stress that the  theory is general enough to apply to any OSC system.  
The theoretical results are validated against wave-optics numerical simulations performed with the {\sc Synchrotron Radiation Workshop} (SRW)~\cite{SRW_chubar} program;. The SRW program is based on the principle of physical optics using Fourier-optics methods to simulate wavefront  propagation~\cite{SRW2,SRW3}. The program also include a Li\'enard-Wiechert solver capable of modeling the radiation emitted as an electron propagates in a given magnetostatic field.  SRW was successfully used to model the optical system associated to OSC taking into account the realistic properties of the emitted radiation from the pickup; see Ref.~\cite{Andorf_NIM} for the implementation. Table~\ref{und_params} lists the undulator and electron-beam parameters considered throughout this paper. These values correspond to the planned OSC experiment at IOTA. The spectrum of the undulator radiation appears in Fig.~\ref{fig:undrad} and shows the angle-integrated full-width-half-maximum relative bandwidth to be on the order of $\sim 30\%$. 

\begin{table}[tttt!!!]
	\begin{tabular}{l l l}
		\hline
		\hline
		parameter, symbol & value & unit. \\
		\hline
		undulator parameter, $K$         & 1.038 & -\\
		length,  $L_u$    & 77.4 & cm           \\
		undulator period,  $\lambda_u$    & 11.06 & cm           \\
		number of periods, $N_u$    & 7 & - \\
		on-axis wavelength, $\lambda_o$         & 2.2 & $\mu$m \\
		electron Lorentz factor, $\gamma$    & 195.69 & - \\
		\hline
		\hline
	\end{tabular}
\caption{Undulator parameters for the OSC test planned at the IOTA facility with a 100-MeV electron beam.}
\label{und_params}
\end{table}
\section{Electric Field Amplitude in the Kicker Undulator}
To find the OSC cooling rates we compute the on-axis electric field of the radiation focused to the kicker undulator. Below, if not stated otherwise, we imply that there is no optical amplifier present. As a major simplification we mitigate depth-of-field effects by locating the undulator such that their centers are at a distance $2R_o\gg L_u$ apart where $L_u$ is the length of one undulator (the pickup and kicker undulator are taken to be identical). A single focusing lens is placed symmetrically between the undulators with its focal length chosen to be $R_o/2$. The problem is split in the following way: first we find the electric field on the lens surface in the time domain and expand it in a Fourier series. We then apply a modified version of Kirchoff's diffraction formula for a lens in order to find the field in the kicker before finally using the field amplitude to compute the longitudinal energy kick to the particle.

The electric field at the lens surface is found by using the Li\'enard-Wiechert formula (in the far-field zone)
\begin{equation}
\pmb{E}(r,t)=\frac{e}{c^2}\frac{\big(\pmb{R}-\pmb{\beta}R\big)\big(\pmb{a}\cdot\pmb{R})-\pmb{a}R\big(R-(\pmb{\beta}\cdot\pmb{R})\big)}{\big(R-(\pmb{\beta}\cdot\pmb{R})\big)^3}
\label{LW}
\end{equation}
where $e$ is the particle charge, $\pmb{\beta}=\pmb{v}/c$, $\pmb{R}=\pmb{r}-\pmb{r}'$ is a vector from the point of emitted radiation $\pmb{r}'$ to the observation point $\pmb{r}$ on the lens surface and $\pmb{a}$ is the acceleration. All values are taken at the emitter (retarded) time $t'=t-R_o/c$. \\
The particle velocity components are 
\begin{equation}
\begin{aligned}
&v_x(t')=-c\theta_e\sin(\omega_ut'), \mbox{~and}\\
&v_z(t')=c\bigg(1-\frac{1}{2\gamma^2}-\frac{\theta_e^2}{2}\sin^2(\omega_ut')\bigg), 
\end{aligned}
\label{vx_vz}
\end{equation}
where $\omega_u\equiv ck_u$ is the undulator frequency and $\theta_e=K/\gamma$ is the amplitude of particle angle oscillation with $K\equiv qB_o/k_umc^2$ being the undulator strength parameter. Plugging Eq.~\ref{vx_vz} into Eq.~\ref{LW} and 
expliciting $\pmb R$ in the cartesian-coordinate system, i.e. $\pmb{R}=R_o[\sin(\theta)\cos(\phi) \hat{\pmb x} +\sin(\theta)\sin(\phi)\hat{\pmb y} +\cos(\theta)\hat{\pmb z}] $, yields the horizontal component of the electric field at the lens
\begin{widetext}
	\begin{equation}
	\begin{aligned}
	&E_x(r,t)=4e\omega_u\gamma^4\cos(t'\omega_u) \times
	&\frac{1+\gamma^2[\theta^2(1-2\cos^2(\phi)-2\theta\theta_e\sin{(t'\omega_u)}\cos(\phi)-\theta_e^2\sin(t'\omega_u)]}
	{cR_o\bigg(1+\gamma^2[\theta^2+2\theta\theta_e\sin(t'\omega_u)\cos(\phi)+\theta_e^2\sin^2(t'\omega_u)]\bigg)}
	\end{aligned}
	\label{Ex1}
	\end{equation} 
\end{widetext}
for angles $\theta \ll 1$.  

 Additionally, and given that only the first harmonic interacts resonantly with the particle in the kicker, we ignore the effect of higher harmonics and relate the observation angle to the radiation frequency via 
\begin{equation}
\omega(\theta)=\frac{2\gamma^2\omega_u}{1+\gamma^2\big(\theta^2+\theta_e^2/2\big)}.
\end{equation}
To find the electric field in the kicker where the radiation has been focused we use a modified Kirchoff's formula
\begin{equation}
E(\pmb{r}'')=\frac{1}{2\pi ic}\int_\Sigma \frac{\omega(\theta)E_w(\pmb{r})}{\abs{\pmb{r}-\pmb{r}''}}e^{i\omega(\theta)\abs{\pmb{r}-\pmb{r}''}/c}d\Sigma
\label{kirchoff}
\end{equation}
where $E_\omega(\pmb{r})$ is the complex amplitude of the first harmonic at a given point on the lens surface
\begin{equation}
E_\omega(\pmb{r})=\frac{\omega(\theta)}{\pi}\int_{0}^{2\pi/\omega(\theta)}E_x(r,t)e^{-i\omega(\theta)t}dt.
\label{fft}
\end{equation}
The integration domain $\Sigma$  in Eq.~\ref{kirchoff} is the surface of the lens where the vector $\pmb{r}$ is located and $\pmb{r}''$ is the coordinate of observation in the kicker. In the absence of dispersion in the lens a ray leaving the pickup with an angle $\theta$ will have a path lengthening equal to $2(R_o\theta^2/2)$ which is exactly compensated by a decrease in the glass thickness of the lens such that all rays take the same amount of time to travel from pickup to kicker centers~\footnote{Taking into account that the depth of field is suppressed it also means that all rays have the same delay from travel between radiating and receiving points in the course of particle motion in the undulators.}. This implies that the argument in the exponential of Eq.~\ref{kirchoff} reduces to a complex constant and can be dropped.
Upon integrating  Eq.~\ref{vx_vz} one obtains an expression for $R(t')$ which can be rearranged to give the observer time as a function of the emitter time
\begin{equation}
\begin{aligned}
t(t')=&\frac{R_o}{c}\bigg(1+\frac{\theta^2}{2}\bigg)+\frac{t'}{2\gamma^2}\bigg(1+\gamma^2\big(\theta^2+\frac{\theta_e^2}{2}\big)\bigg)\\
&-\frac{\theta_e^2}{8\omega_u}\sin(2\omega_ut')-\frac{\theta\theta_e}{\omega_u}\cos(\phi)\cos(\omega_ut').
\end{aligned}
\label{t(p)}
\end{equation}
Computing the derivative $dt/dt'$ together with using  Eqs.~\ref{Ex1} and~\ref{t(p)} enables integration of Eq.~\ref{fft}.
\subsection{Small K parameter}
For the case when $K\ll 1$ Eq.~\ref{Ex1} simplifies to
\begin{equation}
E_x=\cos(\omega_ut')\frac{4e\gamma^4\omega_u\theta_e}{cR}\frac{1+(\gamma\theta)^2(1-\cos^2(\phi))}{\big(1+(\gamma\theta)^2\big)^3}
\end{equation}
Accounting that $\omega(\theta)\approx 2\gamma^2\omega_u/(1+(\gamma\theta)^2)$ and $t(t')\approx t'(1+(\gamma\theta)^2)/2\gamma^2$ yields a straightforward evaluation of Eq. \ref{fft}. \\
Plugging the obtained expression into Eq.~\ref{kirchoff} noting that $\abs{\pmb{r}-\pmb{r}''}\approx R_o$ and assuming a circular lens aperture so that $d\Sigma\approx R_o^2\theta d\theta d\phi$ we find the electric field at the center of the kicker
\begin{equation}
E_x=\frac{4e\gamma^3\omega_u^2K}{3c^2}f_L(\gamma \theta_m),
\end{equation}
where $f_L(x)=1-1/(1+x^2)^3$ described the field reduction due to the finite angular acceptance, and $\theta_m$ is the angle subtended by the lens. \\
Since the depth of field has been suppressed by the large distance between the undulators and lens the amplitude of the electric field becomes constant over the length of the kicker. Longitudinally the field has the dependence $\cos(\omega_ot-k_oz-\psi)$ where $\psi$ is determined by the difference in time of flight between the particle and its radiation. For small $K$ values $z\approx ct(1-1/2\gamma^2)$ and hence the field seen by the electron is $E_x\cos(\omega_ut-\psi)$. Using $d\mathcal{E}/dt=ev_xE_x$ yields the energy transfer to the particle
\begin{equation}
\Delta \mathcal{E}=\frac{2}{3}(e\gamma K k_u)^2L_uf_L(\gamma\theta_m)\sin(\psi).
\label{small_K_Energy}
\end{equation}

\begin{figure}[ttt!!!!!!!!!]
	\centering
	\includegraphics*[width=0.485\textwidth]{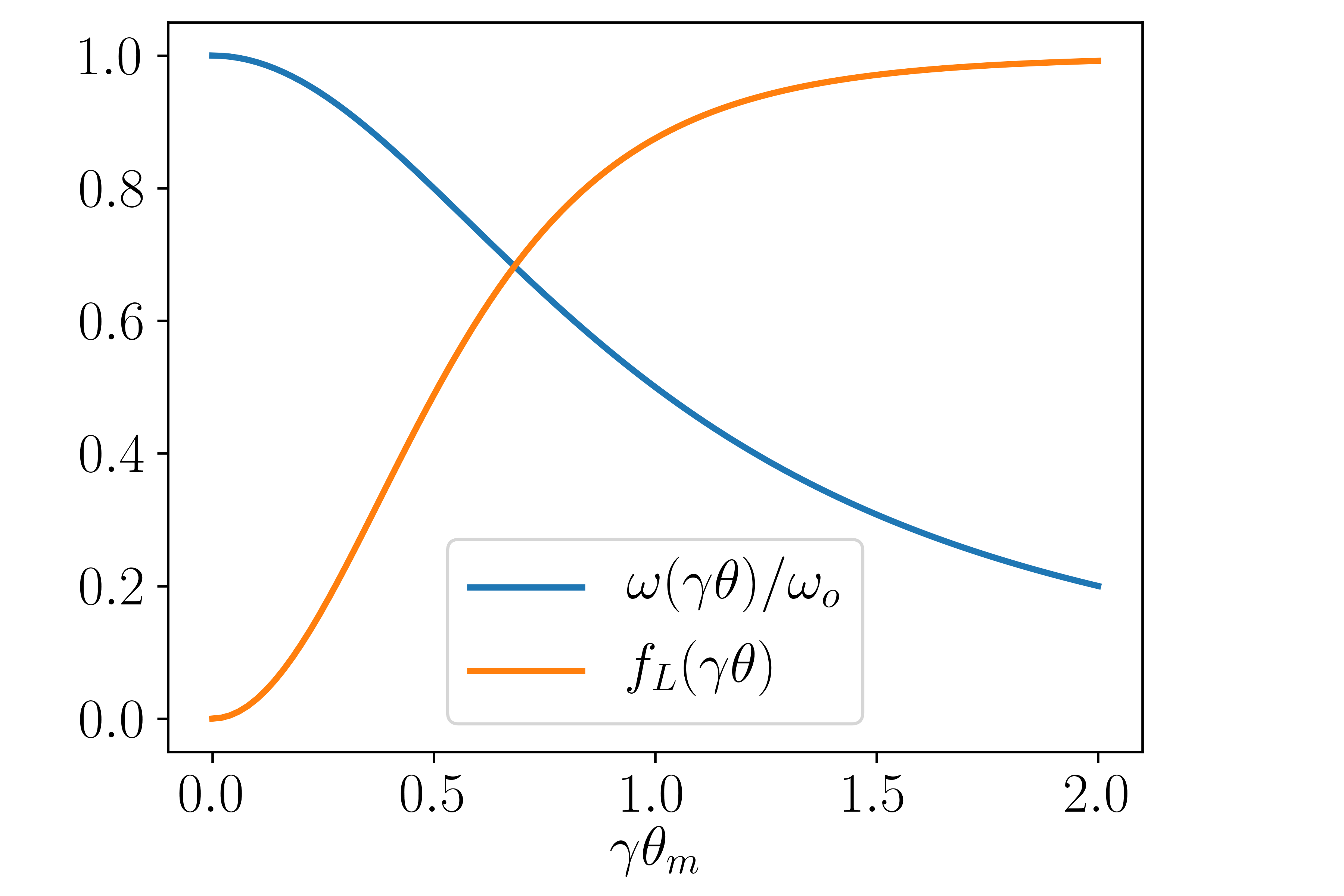}
	\caption{The relative cooling rate as dependent on the angular acceptance of the focusing lens (orange trace) and ratio of forward to outermost angular frequency of the beam (blue trace).}
	\label{fig_angular_BW}
\end{figure}

Figure \ref{fig_angular_BW} shows the relative kick strength as a function of angular acceptance of the focusing lens. We see that an acceptance of $\gamma\theta_m=0.8$ results in approximately 80$\%$ of the theoretical maximum kick. This corresponds to a relative bandwidth acceptance $\Delta \omega/\omega_o$ of 40 $\%$. Here the bandwidth is independent of the number of undulator periods and extends beyond the full-width-half-maximum value. To obtain the actual bandwidth one has to account a finite number of undulator periods. For sufficiently large number of periods this additional spectrum widening can be neglected. \\

In the limit $\gamma\theta_m\gg 1$, Eq.~\ref{small_K_Energy} gives an energy transfer equal to the total amount of energy loss of the particle passing through both undulators in the absence of OSC, i.e. $\Delta \mathcal{E}=\Delta \mathcal{E}_{tot}\equiv \frac{2L_u}{3}(e\gamma K k_u)^2$. If a particle is longitudinally displaced by a distance $s$ relative to the reference particle,  in the course of its travel from pickup to kicker centers then its energy loss is modulated as
\begin{equation}
\Delta \mathcal{E}(s)=\Delta \mathcal{E}_{tot}[1+\sqrt{G}f_L(\gamma\theta_m)\sin(k_os)], 
\label{interference}
\end{equation}
where $k_o=2\gamma^2\omega_u/c$ is the radiation wave number, $G$ the gain of optical amplifier (in power), and the additive 1 in the square brackets accounts for the energy loss in both undulators in the absence of interference, i.e. in the absence of OSC. \\

To find the transverse electric field distribution in the plane orthogonal to the axis and coming through the focal point we must reintroduce the phase term from Eq.~\ref{kirchoff} accounting the phase advance correction. Let the observation point be $\pmb{P''}=(R_o,\rho''\theta'',\rho''\phi'')$ then, Eq.~\ref{kirchoff} becomes 
\begin{widetext}
	\begin{equation}
	\begin{aligned}
	&E_x(\rho'',\phi'')=\frac{8e\gamma^6\omega_u^2\theta_e}{c^2}\int_{0}^{2\pi}\frac{d\phi}{2\pi} \times
	&\int_{0}^{\theta_m}\frac{1+(\gamma\theta)^2(1-2\cos^2\phi)}{\big(1+(\gamma\theta)^2\big)^4}\exp\bigg(\frac{i\rho'k_o\theta\cos(\phi''-\phi)}{1+(\gamma\theta)^2}\bigg)\theta d\theta
	\end{aligned}
	\end{equation}
\end{widetext}
Performing the integration over $\phi$ yields
\begin{equation}
\begin{aligned}
&E_x(\rho'',\phi'')=\frac{8e\gamma^6\theta_e}{c^2}\int_{0}^{\theta_m}\bigg(J_0\bigg(\frac{\rho k_o\theta}{1+(\gamma\theta)^2}\bigg)\\
&+(\gamma\theta)^2J_2\bigg(\frac{\rho k_o\theta}{1+(\gamma\theta)^2}\cos(2\phi'')\bigg)\bigg)\frac{\theta d\theta}{\big(1+(\gamma\theta)^2\big)^4}.
\end{aligned}
\label{trans_E}
\end{equation}

\begin{figure}
	\centering
	\includegraphics*[width=0.485\textwidth]{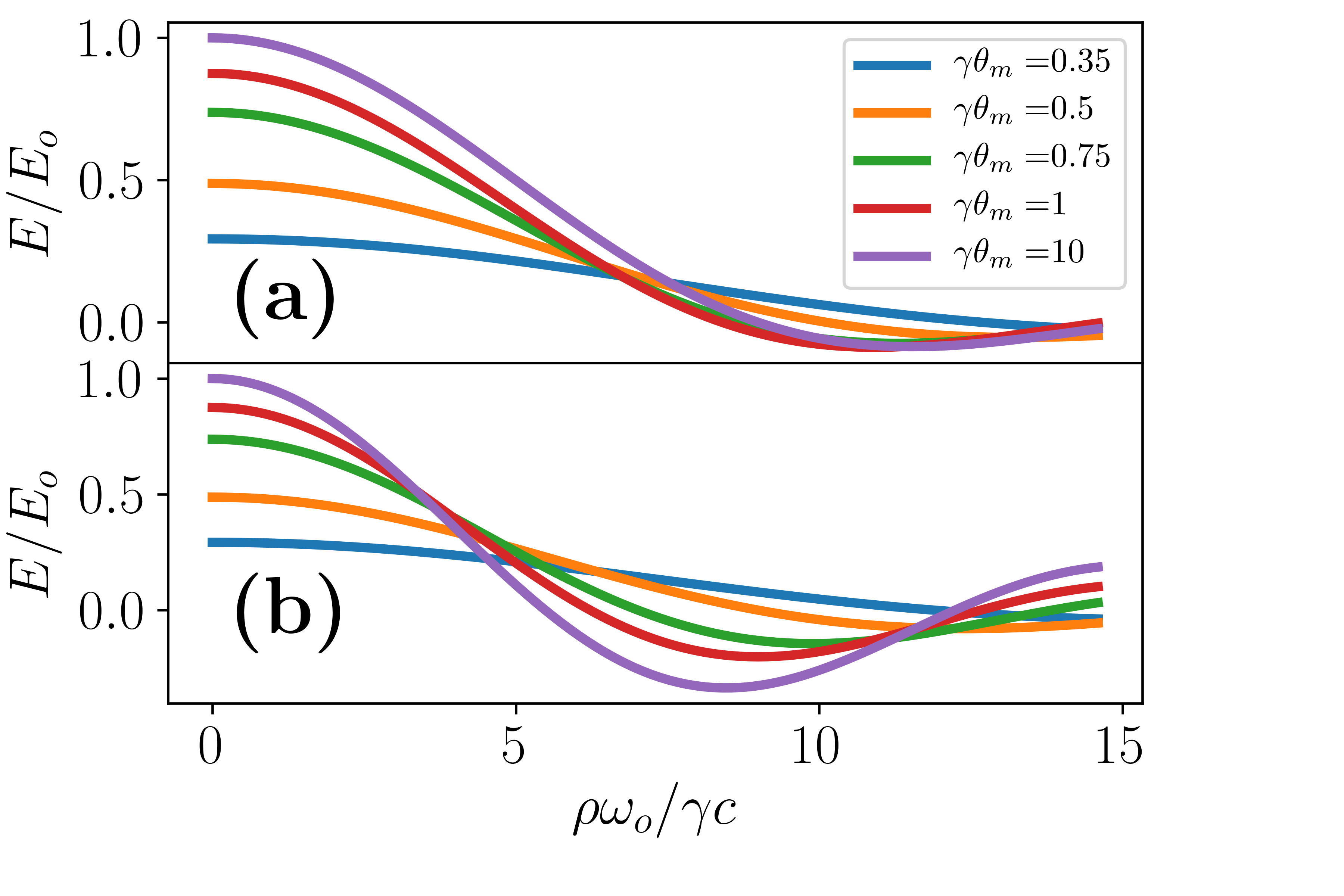}
	\caption{Transverse dependence of the electric field at the focal point in the horizontal $\phi''=0$ (a) the vertical  $\phi''=\pi/2$ (b) plane. The field is normalized to its on-axis value $E_0$. }
	\label{E_norm}
\end{figure} 

Figure \ref{E_norm} shows the relative field dependence in the transverse plane normalized to the limiting case when $\gamma\theta_m=\infty$. \\
The point where the field reaches its first zero determines the half-size for the corresponding plane. If $\gamma\theta_m\ge 0.1$ the horizontal and vertical half-sizes can respectively be estimated as
\begin{equation}
\begin{aligned}
x_o\approx &\lambda_o\sqrt[3]{(1.51\gamma)^3+(0.159+0.619/\theta_m)^3}\mbox{~and,}\\
y_o\approx &\lambda_o\sqrt[3]{(1.08\gamma)^3+(0.619/\theta_m)^3}.
\end{aligned}
\label{x_o_yo}
\end{equation}
\subsection{Arbitrary K value}
\begin{figure}[bbb!!!!]
	\centering
	\includegraphics*[width=0.45\textwidth]{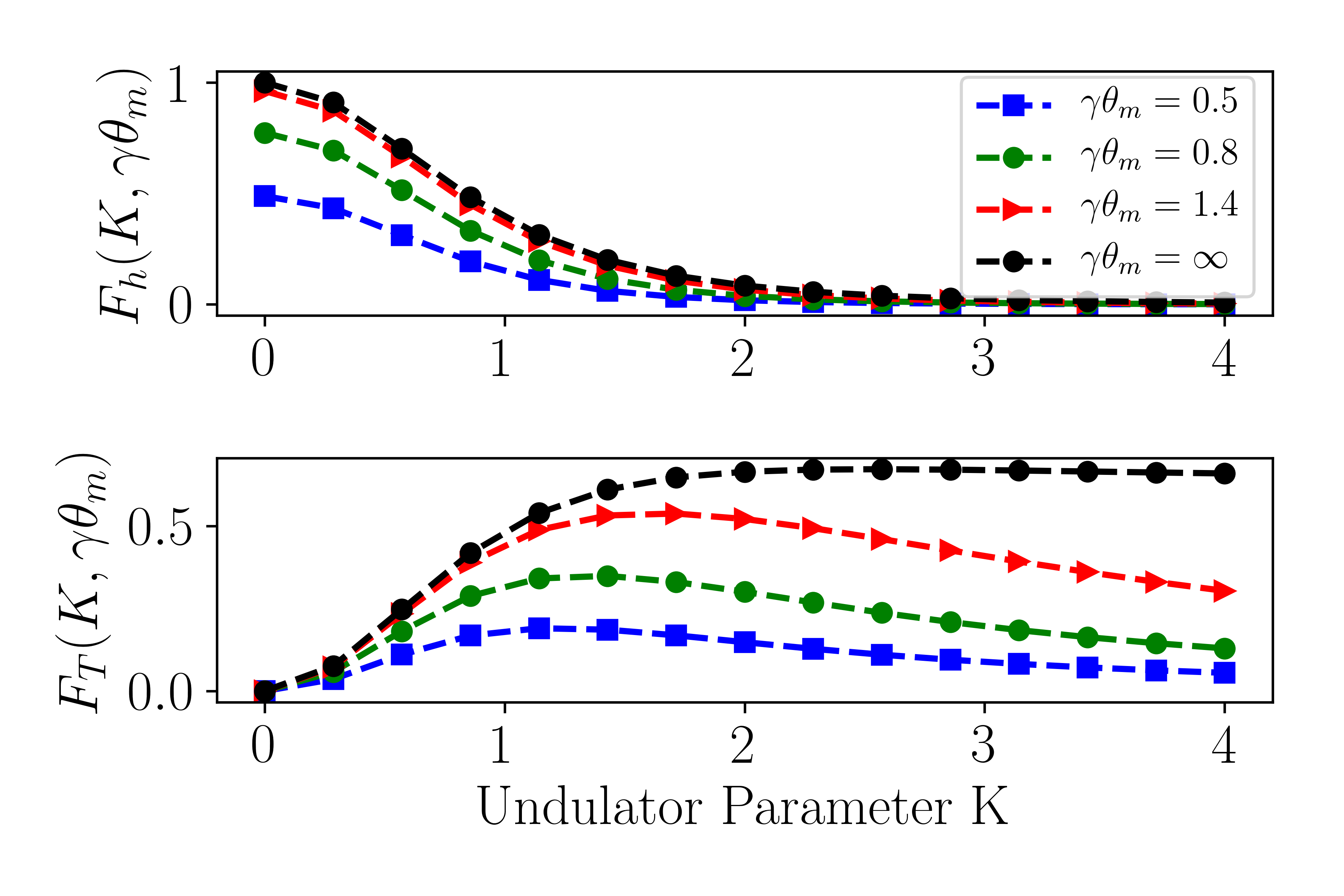}
	\caption{Suppression factors $F_h(K,\gamma\theta_m)$ and $F_T(K,\gamma\theta_m)$ as a function of the undulator parameter $K$ for a range of lens angular acceptances. $K$ is varied while keeping the undulator length and radiation wavelength fixed to respectively $L_u=77.4$~cm and $\lambda_o=2.2$~$\mu$m.}
	\label{kick_fig}
\end{figure}

We now consider the case when $K$ takes an arbitrary value and restrict our attention to the on-axis field. In this case 
\begin{equation}
E_x=\frac{4e\omega_u^2\gamma^3K}{3c^2}F_h(K,\gamma \theta_m)
\end{equation}
where $0\le F_h(K,\gamma \theta_m)\le 1$ is defined in Appendix A and can be evaluated numerically. Integrating along the kicker as before but now also accounting for the longitudinal oscillations of the particle (see Appendix A) yields the OSC kick
\begin{equation}
\Delta \mathcal{E}=\frac{2\pi}{3}e^2k_oN_uF_T(K,\gamma\theta_m)
\label{kick_any_K}
\end{equation}
where $F_T(K,\gamma\theta_m)=K^2(1+K^2/2)F_h(K,\gamma \theta_m)F_u(\kappa)$ is the dimensionless kick amplitude per undulator period. Figure \ref{kick_fig} shows $F_h(K,\gamma \theta_m)$ and $F_T(K,\gamma)$ as a function of $K$ for various angular acceptances.\\

Next, we consider how these formulas are guiding the choice in undulator parameters. Equation~\ref{kick_any_K} indicates that the kick amplitude increases with shorter wavelength. On the other hand, the wavelength must be chosen sufficiently long so that the high amplitude particles are not displaced too far in phase resulting in them being anti-damped.  For proton or heavy ion colliders, the cooling range should be $\sim 5$-$\sigma$ of the beam equilibrium size prior to cooling. Thus the wavelength is chosen as short as possible while still satisfying the required cooling range. An additional constraint dictating the wavelength choice arise when considering the optical amplifier (OA). The addition of the OA introduces additional optical delay over passive OSC. At fixed wavelengths, the cooling range decreases inversely proportional with increasing optical delay~\cite{OSC_val} suggesting that OA gain media operating longer wavelengths are favored. \\

Using a telescope which corrects the depth-of-field effect (see Section~\ref{sec:telescope}) makes the cooling force grow linear with undulator length and thus should be made as long as possible for the given allocated space for a cooling insertion. This leaves $K$ or equivalently $N_u$ to be varied as a free parameter. From the bottom pane in Figure~\ref{kick_fig} we see that for finite lens acceptances growth of $F_T(K,\gamma\theta_m)$ saturates at moderate values of the undulator parameter ($K\le 2$). At even larger $K$ values $F_T(K,\gamma\theta_m)$ begins to decrease but for a fixed undulator length, this reduction is offset by an increased number of periods leading to a leveling off of the kick amplitude for large $K$. \\
\begin{table}[htb]
	\centering
	\begin{tabular}{l c l}
		\hline
		parameter, symbol & value & units \\
		\hline
		peak electric field, $E_x$ & 11.8 & V/m \\
		suppresion factor, $F_h(1.03,0.8)$ & 0.24 &-\\
		period-normalized kick, $F_u(0.17)$ & 0.91 &-\\
		maximum energy exchange, $\Delta \mathcal{E}$ (max) & 22 & meV \\
		electric field half-sizes, $x_o/y_o$ & 680/520 & $\mu$m\\
		\hline
	\end{tabular}
	\caption{Obtained values for OSC using the undulator parameters from Table \ref{und_params}.}
	\label{param_Efield}
\end{table}
For a proof-of-principle demonstration of the OSC using medium-energy electrons \cite{IOTA2} a large $K$ has the detrimental effect of spoiling the equilibrium beam emittance (prior to cooling) since dispersion at both the pickup and kicker is needed for horizontal cooling. Low equilibrium emittance is desirable as it enables OSC to be performed at a short wavelength while still having acceptable cooling ranges and a reasonable optical delay to accommodate an OA. Based on these considerations a value of $K=1.03$ was selected for the OSC experiment in IOTA. Table~\ref{param_Efield} compiles values related to the electric field and kick amplitude for the undulator parameters given in Tab.~\ref{und_params}.
\section{Optical imaging with a Telescope \label{sec:telescope}}
The above results were obtained for the condition that $R_o\gg L_u$. Under this assumption light emitted at a specific longitudinal location in the pickup is nominally refocused to the corresponding location in the kicker, and thus the field amplitude may be considered constant along the length of the kicker. 

In an accelerator, the condition $R_o\gg L_u$ cannot be practically achieved. Instead, a imaging system with a transfer matrix $\mathbf{M}_T$ from pickup to kicker centers equal to $\pm \mathbf{I}$ where $\mathbf{I}$ is the identity matrix ist used.  In this case, the transfer matrix between emitting and receiving points is $\mathbf{O}(l)\mathbf{M}_T\mathbf{O}(-l)=\pm\mathbf{I}$ where $\mathbf{O}(l)$ is the transfer matrix of a drift and $l$ is a displacement measured from pickup/kicker center. \\

The simplest telescope for the $+\mathbf{I}$ case consists of three lenses with one located at the mid-point of the optical transport (lens 2) and two identical lenses located on each side (refer to as lenses 1).  The focal lengths of the lenses 1 and 2 are respectively given by
\begin{equation}
F_1=\frac{L_1L_2}{L_1+L_2}\quad \quad F_2=\frac{L_2^2}{2(L_1+L_2)}
\end{equation} 
where $2(L_1+L_2)$ is the distance from pickup to kicker centers, $F_2$ is placed at the midpoint of the cooling insertion and the $F_1$ lenses are placed on both sides of $F_2$ at a distance $L_2$ away. 
In the case of the transfer matrix equal to $-\mathbf{I}$ we have
\begin{equation}
F_1=L_2 \quad \quad F_2=-\frac{L_2^2}{2(L_1-L_2)}.
\label{MI}
\end{equation}
For a passive OSC test in IOTA the -I telescope matrix is chosen~\footnote{ Such choice is supported by smaller focusing chromaticity and also results in  smaller transverse separation between radiation and particle in the kicker 
undulator. For chosen IOTA beam optics this condition is fully satisfied for horizontal plane while only partially for the vertical plane.} and the lens focal lengths and positions are given in Tab.~\ref{param_telescope}.

For the active test, the telescope has the additional requirement of tight focusing of the pickup radiation in the amplifier. For this case, the $+\mathbf{I}$ telescope must be used. In passing, we note that by setting $L_1=L_2$ in Eq. \ref{MI} the center lens can be eliminated. This telescope would be desirable to use as the weaker focusing lenses would reduce chromatic effects. It, however, cannot be used in IOTA as it would place the lenses in the path of the particle beam.

The performances of the telescope was studied using SRW \cite{Andorf_NIM}. As expected the telescope supports longitudinal point-to-point imaging along the kicker as evident from the time-domain waveform modulation. A reduction in field amplitude occurs at the edges of the kicker as the effective angular aperture is reduced for light emitted at the pickup edges. For parameters anticipated in IOTA this effect reduced the kick amplitude by $10\%$.
\begin{table}[htb]
	\begin{tabular}{l c c l}
		\hline
		parameter & symbol & value & units \\
		\hline
		distance from kicker to lens 1 & $L_1$ & 143 & cm \\
		distance from lens 1 to lens 2 &$L_2$ & 32 &cm\\
		focal length of lens 1 &$F_1$ & 32 &cm\\
		focal length of lens 2 &$F_2$ & 4.6 & cm \\
		\hline
	\end{tabular}
	\caption{Geometrical parameters of lens telescope for the passive-OSC proof-of-principle experiment in IOTA.}
	\label{param_telescope}
\end{table}
\section{Focusing Errors and Lens chromaticity}
We now consider two sources of errors: (1) manufacturing errors of the telescope lenses, and (2) lens chromaticity which plays a significant role due to the wide bandwidth required for the OSC. To simplify the problem we consider these errors separately. 
\subsection{Focal length error}
\begin{figure}[h]
	\centering
	\includegraphics*[width=0.45\textwidth]{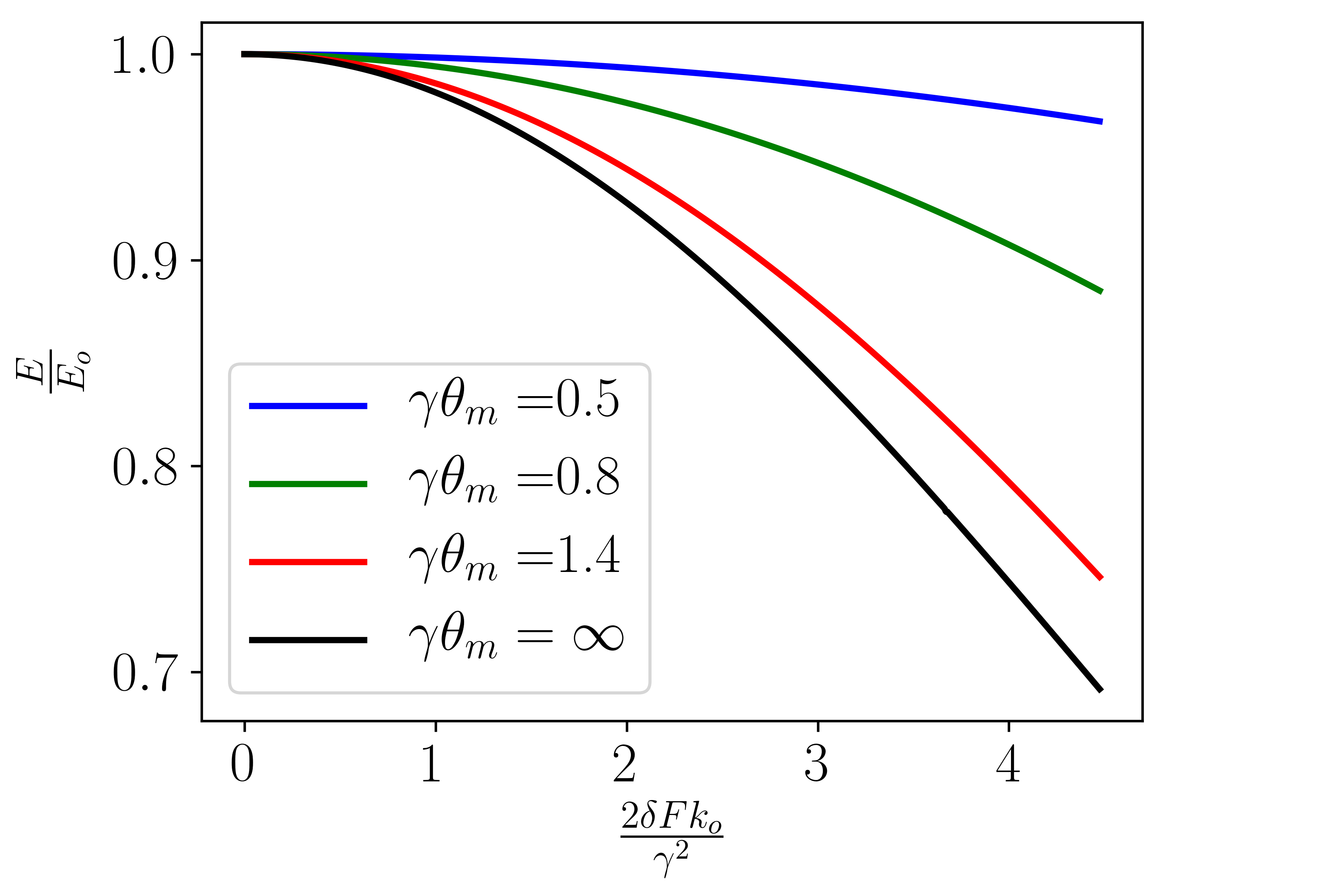}
	\caption{Relative reduction of the on-axis electric field at the focal point as a function of the dimensionless focusing error and for different angular acceptances of lens telescope.}
	\label{fig_focal_error}
\end{figure}

We first consider the case of a single focusing lens and assume $K\ll 1$ in which case the electric field at the focus becomes:
\begin{equation}
E_x=\frac{8e\gamma^5\omega_u^2K}{c^2}\abs{\int_{0}^{\theta_m}\exp\big(i\Phi(\theta)\big)\frac{\theta d \theta}{\big(1+(\gamma\theta)^2\big)^4}}
\end{equation}
where $\Phi(\theta)$ is an additional phase term accounting the error in focusing strength. The phase advance of light passing through the lens will be given by $k(\theta)\rho^2/(2(F+\delta f))$. Then since, at the lens surface, $\rho=2F\theta$ and assuming $\delta F\ll F$ we find $\Phi(\theta)=-2\theta^2k_o \delta F/(1+(\gamma\theta)^2)$. Figure~\ref{fig_focal_error} shows the relative reduction in the field at the focal point as a function of $\delta_F$ for a variety telescope angular acceptances. Taking the case of $\gamma\theta_m=0.8$ and requiring the focusing errors not to reduce the field by more than $2\%$  requires $\delta F<1.23$ cm.\\
To find tolerances of the focusing errors of the telescope lenses  we, first, consider a single lens telescope. A focusing error of $\delta F$ results in a displacement of focal point $\delta s = 4\delta F$. Then, we require that a focusing error for a single lens of the telescope results in the same displacement of the focal point as in the single lens case. Introducing relative errors of telescope lenses ($f_k\equiv-\delta F_k/F_k$, $k=1,2,3$ ) multiplying the corresponding matrices and leaving only the first order terms we obtain for the $M_{12}$ transfer matrix element:
\begin{equation}
M_{12}=-2(L_1-L_2)f_2+\frac{L_1^2}{L_2}(f_1+f_3).
\end{equation}
The focal point displacement is equal to $-M_{12}/M_{22}$. Accounting that the matrix is close to identity matrix ($\abs{M_{22}}\approx 1$) one obtains $\delta s\approx M_{12}$. Considering only one focal length error at a time, we find that for the telescope parameters presented in Tab.~\ref{param_telescope} that the fractional error on the focal length value should be within $\abs{f_1}=\abs{f_3}\le0.8\%$ and $\abs{f_2}\le2.3 \%$. Performing similar calculations for the displacement of the outer lenses one obtains $\abs{\delta L_2}\le2.3$ mm. Note that longitudinal displacement of the entire telescope results in no error because of its point-to-point imaging property. 
\subsection{Lens Chromaticity}
One of the main challenges with the optical system comes from the large relative bandwidth $\sim 30 \%$ associated to the radiation; see Fig.~\ref{fig:undrad}. Chromaticity within the lenses results in mis-focusing of the radiation in the kicker. The lens focal length depends on wavelength as $F=F_o(n_{opt}-1)/(n(\lambda)-1)$, where $n_{opt}$ is index of refraction at $\lambda_{opt}$, the wavelength that minimizes the telescope focusing errors. In the case of the OSC experiment at IOTA, Barium fluoride (BaF$_2$) was selected for the lens material owing to its low chromaticity over the OSC bandwidth [2.2, 3.2]~$\mu$m. The wavelength dependence of the index of refraction is described by the Sellmeier's equation~\cite{Refractive_index_HLI}
\begin{eqnarray}
n(\lambda)^2 &=&1.33973+\frac{0.81070\lambda^2}{\lambda^2-0.10065^2}+ \nonumber \\
&& \frac{0.19652\lambda^2}{\lambda^2-29.87^2}+\frac{4.52469\lambda^2}{\lambda^2-53.82^2}.
\end{eqnarray}
For $\lambda=2.2$~$\mu$m we have $dn/d\lambda=-3.21\times 10^{-3}$~$\mu$m$^{-1}$ and a group-velocity dispersion (GVD) value of -9.7405 fs$^2$/mm. The optical-transport transfer matrix from pickup to kicker can now be computed as a function of wavelength and the displacement of the focal point as a function of the wavelength can be inferred using $\delta s(\lambda)=-M_{12}(\lambda)/M_{22}(\lambda)$. Intuitively $\lambda_{opt}$ will be somewhere in the middle of the OSC band where the photon flux and the growth rate of the kick amplitude with respect to lens angular acceptance are largest. This leads us to choose $\lambda_{opt}=2.6$ $\mu$m. Computation of $\delta s(\lambda)$ results in a focal displacement ranging from 3 to -4 cm within the interval $\lambda \in [ 2.2, 3.2]$~$\mu$m.  The increase of the total spot size at the focal point is estimated as 
\begin{equation}
r(\lambda)=\sqrt{\frac{x_o^2+y_o^2}{2}+\big(\theta_m\delta s (\lambda)\big)^2}, 
\end{equation}  
where $x_o$ and $y_o$ were computed using Eq.~\ref{x_o_yo}. It results in a 2$\%$ and $4\%$ increase in the spot size at 2.2 and 3.2 $\mu$m. As a first order approximation the field (and hence the kick amplitude) is reducing by $1/r$ and thus we expect focusing chromaticity to reduce the OSC kick by at most of few percent in IOTA. 

\begin{figure}[h]
	\centering
	\includegraphics*[width=0.485\textwidth]{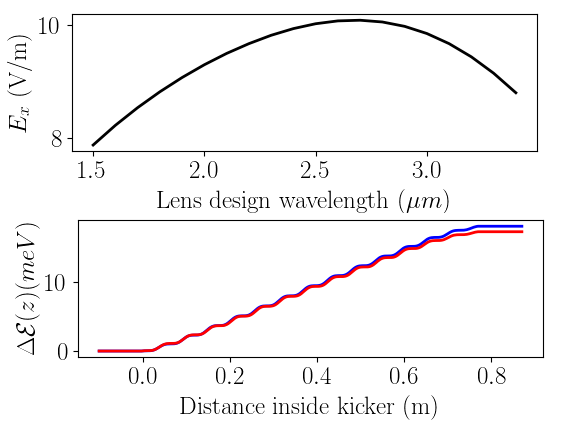}
	\caption{Top: The electric field amplitude at the kicker center for different design wavelengths of the telescope lenses. Bottom: The energy exchange for a particle copropagating with the simulated radiation field from SRW through the kicker  with (red trace) and without (blue trace) accounting for the lens dispersion.}
	\label{fig_chroma}
\end{figure}

To confirm our choice of $\lambda_{opt}$ the on-axis field was computed with SRW for a range of design wavelengths for the lenses. Since SRW computations are done in the Fourier domain it is straightforward to account for lens dispersion. 
The top panel in Fig.~\ref{fig_chroma} shows the resulting dependence of electric-field amplitude on the design wavelength of the telescope. The bottom pane displays the energy gain for an electron with its arrival phase chosen to be $\psi=\pi/2$ with and without accounting for dispersion. Including dispersive effects decreases the kick by approximately $4\%$. The SRW simulatiozns also include pulse broadening (reduction in field amplitude) from the second-order dispersion. 
\section{Conclusion}
The presented semi-analytic approach allows one to compute the energy kick a particle receives in a single pass through an OSC insertion. It provides a straight forward way of determining the main undulator and optical telescope parameters and tolerances. 
Above we mostly considered a passive scheme for the OSC which is suitable only for a proof-of-principal demonstration with electrons. For cooling of hadrons an amplifier providing 20-30~dB within a single pass is required. As a first order estimate the kick amplitude is proportional to $\sqrt{G}$ of the pickup radiation gain in power. A further refinement that accounts the finite amplifier bandwidth (and possibly) additional 2nd order dispersion will reduce the overall kick value. These effects are easily accounted with SRW. For example in an active test of the OSC in IOTA an amplifier, based on Cr:ZnSe, gives 7~dB of gain. However the kick amplitude is expected to increase only by a factor 1.65 when host dispersion and amplifier bandwidth are taken into account~\cite{AndorfOSA}.

Throughout our analysis, we have considered radiation emitted by a single particle. However, in practice, a given particle will be affected by the radiation emitted by its neighboring particle in addition to its own. Such an interparticle interaction will result in a phase-space "heating" effect which is straightforward to describe using the theory developed for conventional stochastic cooling~\cite{Mohl}. Neglecting this heating effect in the present analysis is justified as the considered proof-of-principle experiment at the IOTA ring involves passive OSC or low-gain amplification and a low number of particle in the OSC sampling slice.  \\

The model presented in this paper was implemented in the {\sc elegant}~\cite{elegant} program to successfully verify, via particle tracking, the main features of OSC cooling for the experiment planned at IOTA~\cite{Matt_ELEG}. We finally note that the formalism presented in this paper is general and could be extended to other OSC configurations. It could additionally prove useful in other electron-radiation interaction mechanisms involving the transport and manipulation of radiation from undulator magnets.

\section{Acknowledgments} 
The authors would like to express their gratitude to J. Ruan and A. Romanov from Fermilab. The work of M.A. and P.P. was supported by the US Department of
Energy under contract DE-SC0013761 to Northern Illinois University. Fermilab is managed by the Fermi Research Alliance, LLC for the U.S. Department of Energy
Office of Science Contract number DE-AC02-07CH11359.
\appendix
\section{Electric field \& kick amplitude for arbitrary values of $K$}
In this appendix we derive an expressions for the electric field and kick amplitude for arbitrary values of $K$. We begin with Eq.~\ref{fft} with the integration variable changed from emitter to observer time
\begin{equation}
E_\omega(\theta,\phi)=\frac{\omega(\theta)}{\pi}\int_{0}^{\frac{2\pi}{\omega_u}}E_x(t')\exp\big(-i\omega(\theta)t(t')\big)\frac{dt}{dt'}dt'.
\end{equation}
Insertion of Eq.~\ref{Ex1}, Eq.~\ref{t(p)} and its derivative into the above equation and performing some simplifications yields
\begin{widetext}
	\begin{equation}
	\begin{aligned}
	&E_\omega(\theta,\phi)=\frac{4\gamma^4e\omega_u^2\theta_e}{\pi c R_o}
	\int_{0}^{2\pi}\bigg[F_c(\Theta,K,\tau',\phi)\times 
	&\frac{1+\Theta\big(1-2\cos^2(\phi)\big) -2\Theta K \cos (\phi)\sin(\tau')-K^2\sin^2(\tau')}{\big[1+\Theta^2 +2\Theta K\cos(\phi)\sin(\tau')+K^2\sin^2(\tau')\big]^3}\bigg]d \tau'
	\end{aligned}
	\end{equation}
	where
	\begin{equation}
	\begin{aligned}
	&F_c(\Theta,K,\tau',\phi)=\cos(\tau')\times
	&\exp\big[-i\tau'+i\frac{K^2\sin(2\tau')+8\Theta K \cos(\phi)\cos(\tau')}{4\big(1+\Theta^2+K^2/2\big)}\big]\times
	&\big[1+\frac{4\Theta K\cos(\phi)\sin(\tau')-K^2\cos(2\tau')}{2\big(1+\Theta^2+K^2/2\big)}\big]
	\end{aligned}
	\end{equation}
\end{widetext}

and the variables $\Theta=\theta\gamma$ and $\tau'=\omega_ut'$ were introduced. This expression can then be inserted into Eq. \ref{kirchoff} which will yield the result
\begin{equation}
E_x=\frac{4\gamma^3e\omega_u^2K}{3 c^2}F_h(K,\gamma\theta)
\end{equation}
where
\begin{widetext}
	\begin{equation}
	\begin{aligned}
	F_h(K,\gamma\theta)=\frac{3}{\pi^2}\int_{0}^{\Theta_m}\int_{0}^{2\pi}\int_{0}^{2\pi}\bigg[\frac{F_c(\Theta,K,\tau',\phi)}{1+\Theta^2+K^2/2} \times
	\frac{1+\Theta\big(1-2\cos^2(\phi)\big) -2\Theta K \cos (\phi)\sin(\tau')-K^2\sin^2(\tau')}{\big[1+\Theta^2 +2\Theta K\cos(\phi)\sin(\tau')+K^2\sin^2(\tau')\big]^3}\bigg]d \Theta d \phi d\tau'
	\end{aligned}
	\end{equation}
\end{widetext}
Next to find the kick we integrate the energy exchange between the particle and field.  For a finite $K$ value, in addition to its transverse motion, the particle also oscillates longitudinally. This will lead to an efficiency factor $F_u(\kappa)\le 1$ that must be included in computation of the kick amplitude. As before we assume the electric field amplitude seen by the electron stays constant as it propagates along with the particle in the kicker. \\
Let the field be described as $E=E_x\cos(\omega_ot-k_oz)$. Integration of $v_z$ given in Eq.~\ref{vx_vz} gives
\begin{equation}
z=c\bigg[\bigg(1-\frac{1}{2\gamma^2}-\frac{\theta_e^2}{4}\bigg)t+\frac{\theta_e^2}{8\omega_u}\sin\big(2\omega_u t)+z_o\bigg].
\end{equation}
Inserting this into the expression for the field, recalling that $\omega_u=\omega_o(1/2\gamma^2+\theta_e^2/4)=\omega_o(1+K^2/2)/2\gamma^2$ and defining $\kappa=\theta_e^2\omega_o/8\omega_u=K^2/(4+2K^2)$ yields:
\begin{equation}
E=E_x\cos(\omega_ut-\kappa\sin(2\omega_ut)-\psi).
\end{equation}
Then using $d\mathcal{E}/dt=ev_xE_x$ gives
\begin{eqnarray}
\Delta \mathcal{E}&=&\frac{ecKE_x}{\gamma}\int_{0}^{\frac{2\pi N_u}{\omega_u}}\cos\big(\omega_u t -\kappa\sin(2\omega_u t)-\psi\big) \nonumber \\
& & \vspace{1cm} \times \sin(\omega_u t)dt.
\end{eqnarray}
The above integral is easily evaluated with the substitution $\xi=\omega_u t$ and the aid of the following two identities 
\begin{equation}
\begin{aligned}
&\int_{0}^{2\pi}\cos\big(\xi-\kappa\sin{2\xi}\big)\sin \xi d\xi=0\\
&\int_{0}^{2\pi}\sin\big(\xi-\kappa\sin{2\xi}\big)\sin \xi d\xi=\pi\big(J_0(\xi)-J_1(\xi)\big)
\end{aligned}
\end{equation}
which yields
\begin{equation}
\Delta \mathcal{E}	=\frac{eKE_xL_u}{2\gamma}\sin(\psi)F_u(k)
\end{equation}
where 
\begin{equation}
F_u(\kappa)=J_0(\kappa)-J_1(\kappa).
\end{equation}
Thus accounting for the longitudinal motion of the particle reduces the kick by a factor $F_u(\kappa)$.\\

\end{document}